# DFT and TDDFT calculation of lead chalcogenide clusters up to (PbX)$_{32}$

V. S. Gurin

*Research Institute for Physical Chemical Problems, Belarusian State University,*
*Leningradskaya 14, 220006, Minsk, Belarus*

**Abstract**

A series of the lead chalcogenide clusters Pb$_n$X$_n$ (X=S,Se; n=4,8,16,32) with structures as fragments of the bulk crystalline lattice are calculated at DFT level with B3LYP functional and ECP basis set. Optical absorption spectra are simulated through the TDDFT method. The results are in consistence with experimental data PbS and PbSe for magic size clusters of this size range.

## 1. Introduction

Nanoscale lead chalcogenides are of great interest last years for potential design of optical devices covering a wide spectral range from visible to infrared [1,2]. Lead chalcogenides (PbX) are narrow-band semiconductors and PbX nanoparticles and nanoclusters reveal the more pronounced quantum-size effects than major II-VI compounds. The strongest changes in properties occur for few-atomic nanoclusters, so called magic size clusters, and they were studied recently both experimentally [3] and theoretically [4] for different examples of their composition.

In this paper, we consider a series of models for (PbX)$_n$ (n=4,8,16,32) clusters built on the basis of crystalline lattice of the rock salt type which exists for the bulk PbS, PbSe, and PbTe. Structures of chalcogenide clusters those have been established experimentally and analyzed theoretically are rather versatile, they include both bulk-like geometries and some arbitrary versions. Stoichiometry of binary compounds, II-VI and IV-VI series (e.g. CdX, ZnX and PbX), in the case of clusters can be also different due to the surface capping by ligands terminating dangling bonds. An overview of complete set of possible cluster structures is the topic of separate reviews, however, for lead chalcogenides the bulk-like stoichiometric clusters built from the fragments of rock salt lattice are rather feasible versions (but, naturally, *a priopi* not of minimum energy). Experimental studies of PbX magic size clusters reveal the crystalline structures close to the rock salt (with modified unit cell parameters).

## 2. Calculation technique

Calculations were done using DFT (density functional theory) with extension to time-dependent version (TDDFT) that allows calculate both geometry of the ground state and electronic transitions to the excited ones of the cluster models with many atoms of heavy elements at reasonable computational resources. In spite of many calculations published to date at different level of theory for various composition of lead chalcogenide clusters, there is noticeable variance in approaches and calculation details (type of functional, effective core potentials (ECP), basis sets, etc.) resulting in slightly different results both for geometrical parameters and electronic structure. Direct verification of geometry is not too easy because problems with extraction and isolation of certain clusters, while optical absorption spectra simulated through electronic transitions between molecular orbitals in clusters can be compared with available experimental data. However, there is also variance within the expectable range of transition energies derived by different calculations. Within the framework of this publication we tested one of common standard choice for functional and basis set with ECP for the elements of these clusters, B3LYP and LANL2DZ, respectively. This functional is known as a successful balance of exchange and correlation contributions, and this ECP basis treats valence electrons with enough accuracy for p-elements, also with partial inclusion of relativistic effects. Gaussian 09 package was used for the calculations [5].

Fig. 1 demonstrates the cluster models $Pb_nX_n$, X=S,Se, n=4,8,16,32 those are fragments of the rock salt crystal lattice inherent for the bulk PbS and PbSe. We keep the cubic symmetry under the calculations.

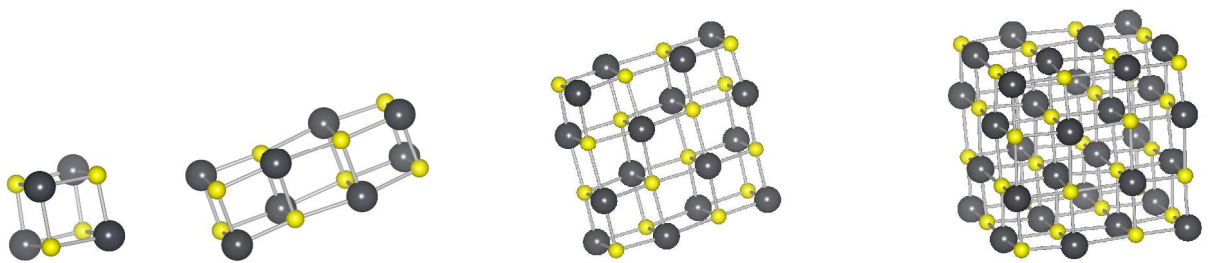

Fig. 1. Structures of the cluster models of $Pb_4X_4$, $Pb_8X_8$, $Pb_{16}X_{16}$, $Pb_{32}X_{32}$, X=S,Se, used for calculations

## 3. Calculation results

Table 1 and Fig. 2 present the principal calculation results for this set of lead chalcogenide clusters. These data show that the clusters are stable and the bonding energy increases with size increase that is quite expectable since the clusters with 64 atoms are closer to the bulk counterparts, while $Pb_4X_4$ include much part of surface atoms. The bond lengths are less than in the bulk compounds, 2.95, and 3.15 Å, respectively, but there is no monotonous dependence with the cluster size. The gap between the frontier orbitals, HOMO-LUMO (that can be considered as a zeroth approximation for the band gap) appears to be significantly less for $Pb_4X_4$ than for remainder $Pb_nX_n$, n=8,16,32. It is noticeable that this value is almost the same for sulfide and selenide clusters.

Table 1. Calculation data for a series of $Pb_nX_n$ clusters (X=S, Se), n=4,8,16,32.

| Clusters | Functional/Basis set | Pb-S/Se bond length, Å | Total electronic, energy, a.e. | Bonding energy of clusters, eV | First lines of abs spectra, nm | ΔE HOMO-LUMO, eV |
|---|---|---|---|---|---|---|
| $Pb_4S_4$ | B3LYP/LANL2DZ | 2.767 | -54.6889 | 33.201/4=8.300 | 397;318;283 | 3.897 |
| $Pb_4Se_4$ | B3LYP/LANL2DZ | 2.903 | -51.1726 | 30.940/4=7.375 | 428;336;316 | 3.632 |
| $Pb_8S_8$ | B3LYP/LANL2DZ | 2.714 | -109.4080 | 67.2245/8=8.403 | 409;405;398 | 5.745 |
| $Pb_8Se_8$ | B3LYP/LANL2DZ | 2.853 | -102.3715 | 62.5958/8=7.824 | 460;455;431 | 5.801 |
| $Pb_{16}S_{16}$ | B3LYP/LANL2DZ | 2.664 | -218.8705 | 135.9239/16=8.495 | 465;456;450 | 5.628 |
| $Pb_{16}Se_{16}$ | B3LYP/LANL2DZ | 2.810 | -204.7913 | 126.5059/16=7.907 | 499;488;477 | 5.765 |
| $Pb_{32}S_{32}$ | B3LYP/LANL2DZ | 2.758 | -437.8101 | 273.7446/32=8.555 | 509;488;477 | 5.682 |
| $Pb_{32}Se_{32}$ | B3LYP/LANL2DZ | 2.892 | -409.6559 | 255.0064/32=7.969 | 534;523;514 | 5.560 |

The absorption spectra for $Pb_{32}X_{32}$ are presented in Fig. 2 as positions of electronic transitions in the wavelength scale indicating relative intensities derived from the corresponding oscillator forces. This series of the spectra falls into the visible and near UV ranges similar to various experimental ones for the magic size clusters PbS and PbSe in the same range of atom numbers. The spectra for $Pb_{32}S_{32}$ are slightly blue shifted with respect to $Pb_{32}Se_{32}$ that is usually for different couples of chalcogenides.

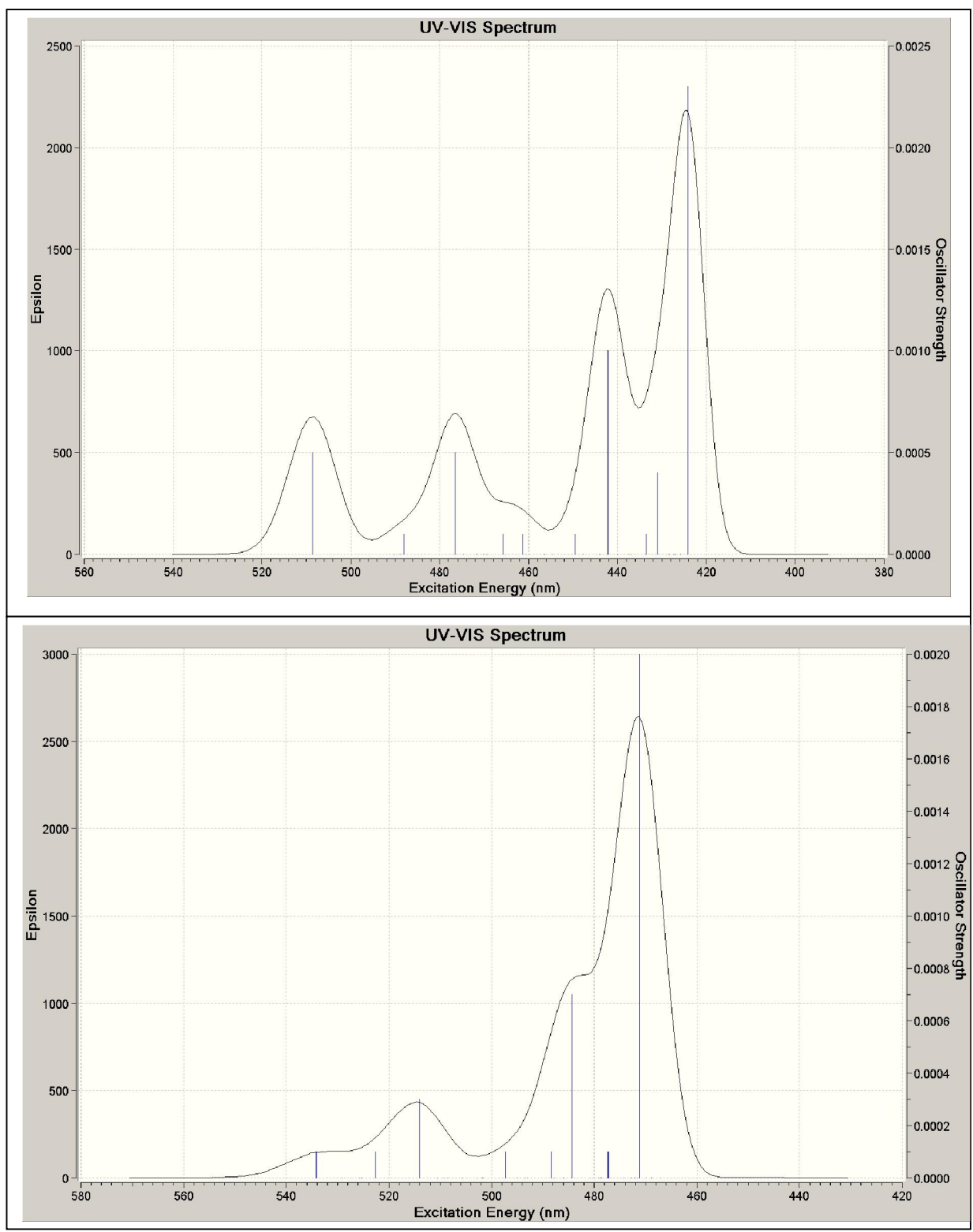

Fig. 2. Absorption spectra calculated by the TDDFT method for $Pb_{32}S_{32}$ (top) and $Pb_{32}Se_{32}$ (bottom) clusters.